*Research Note*

# USE OF NoSQL DATABASE AND VISUALIZATION TECHNIQUES TO ANALYZE MASSIVE SCHOLARLY ARTICLE DATA FROM JOURNALS

Gouri Ginde[1], Snehanshu Saha[2], Archana Mathur[3], Harsha Vamsi[2], Sudeepa Roy Dey[2], Swati Sampatrao Gambhire[2]
[1]Department of Computer Science and Information Systems, University of Calgary, Canada
[2]Department of Computer Science and Engineering, PES University, Bangalore
[3]Indian Statistical Institute, SSIU, Bangalore

**ABSTRACT**

Visualization of the massive data is a challenging endeavour. Extracting data and providing graphical representations can aid in its effective utilization in terms of interpretation and knowledge discovery. Publishing research articles has become a way of life for academicians. The scholarly publications can shape-up the professional growth of authors and also expand the research and technological growth of a country, continent and other demographic regions. Scholarly articles have grown in gigantic numbers that are published in different domains by various journals. Information related to articles, authors, their affiliations, number of citations, country, publisher, references and other information is like a gold mine for statisticians and data analysts. This data when used skilfully, via visual analysis tool, can provide valuable understanding and can aid in deeper exposition for researchers working in domains like scientometrics and bibliometrics. Since the data is not readily available, we used Google scholar, a comprehensive and free repository of scholarly articles, as data source for our study. Data was scraped from Google scholar and stored as a graph and later visualized in the form of nodes and its relationships, which offered discerning and concealed information of growing impact of articles, journals and authors in their domains. Not only this, evident domain shift of an author, various research domains spread for an author, predicting emerging domain and subdomains, detecting cartel behaviour at Journal and author-level was also depicted by graphical analysis. Neo4j graph database was used in the background to help store the data in structured manner.

## INTRODUCTION

Data visualisation is the visual representation of patterns in data. It can help to understand and relate to the data, communicate and represent data in a more comprehensive manner to others. Data visualisation can be as trivial as a simple table, elaborate as a map of geographic data depicting an additional layer in Google Earth or complex as a representation of Facebook's social relationships data. Visualisation can be applied to qualitative as well as quantitative data. Visualisation has turned into an inexorably well-known methodology as the volume and complexity of information available to research scholars has increased. Also, the visual forms of representation have become more credible in scholarly communication. As a result, increasingly more tools are available to support data visualisation.

### Need for effective visualization of scholarly publications

Scholarly articles and journals have always been a subject of interesting research, mainly because of the stakes involved in the scholarly publications primarily in the world of academia. These articles are weighed based on the credibility of the journals first and then the authors. Nevertheless there does not exists one complete solution which can provide complete visualization of the data at author, country, and journal and domain level for all the journals in the world. This data, when explored using right tools, has a tremendous potential to unearth interesting patterns which can expose illegitimate cartel behavior at various levels as well as enunciate out performing under mined authors and evolving journals. Massive scale journals data can be used for the purpose of a data visualization is analysis,

The scibase and scientometric modeling effort is endorsed and supported by IEEE Computer Society Bangalore Chapter.

communication, or both. Analysis requires careful attention to the parameters used. Different parameters reveal different patterns, and it is challenging to determine which are significant with respect to the key research questions.

*Importance of data visualization*
High impact visualization is like a picture speaking a thousand words. Selecting good visual technique to display the data holds key to a good impact. Fancy bubble charts, time domain based motion graphs are possible now because of the languages such as python and R. However, selecting the one which can effectively represent the data to the audience is crucial. To begin with, in this paper we have visualized data using in-built D3.js script available with Neo4j Graph database. Further on we have used the pie chart, line graph and area graphs to represent the information concisely to overcome data overload through dynamic presentation in the initial stage.

Fig 1 shows the flow diagram of the complete system. A huge data from Google scholar has been scraped over a period of 2 Months [1] using web scraping methodology to acquire the required dataset. Further this data is pre-processed and fed into Neo4j graph database using advanced cyphers. These cyphers deconstruct complex JSON documents and quickly turn them into a graph structure of rich relationships without duplication of information. In the next stage the data from the graph database is queried using various cyphers to compute a few more scholastic indicators such as self-citations, total citations, international collaboration ratio etc. These scholastic indicators are then created as the properties at article and journal level in the graph database. Finally, various questions are transformed into cypher queries to get the meaningful data from the graph database. This data is then visualized using various charts and graphs.

*Data curing*
Data accumulation, a first step in data curing, is an arduous task for any research project. For this research Google scholar has been used as resource as it provides a comprehensible and complete data required for good analysis. However, Google scholar does not provide any API. Hence web scraping methodology has been used to gather the data [1, 4]. Next, accumulated data is pre-processed, an intermediate task where the data is cured and made ready for further analysis. The scraped data, which is in JSON format, is first trimmed of any unwanted characters. Then, data is cleansed of Unicode characters.

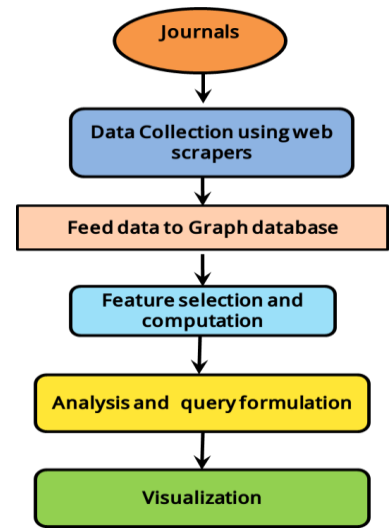

Figure 1. Flow diagram

In order to provide interesting visualization patterns, few of the parameters had to be derived from accumulated data. Pre-processing also involves computation of these scholastic indicators/parameters. The scraped data, which is in JSON format, is in the pure textual form hence, cosine similarity string metric has been used for text comparison in place of pure string comparison operation for better results.

Using cosine similarity string metric, author-level and journal-level self-citation, international collaboration ratio and other scholastic indicators are computed. Self-citation count is a part of citation count when a citing article shares at least one author name with the article it is citing. Journal-level self-citation is part of citation count when a Journal's article cites another article published by the same journal.

*Cobb Douglas model for computation of 'Internationality'*
1. Definition of 'Internationality' of a journal
Internationality has been defined and perceived as the degree to which a journal transcends local communities and boundaries, with respect to the quality of publication and influence. We define internationality of peer-reviewed journals as a measure of influence that spreads across boundaries and attempts to capture different and hitherto unperceived aspects of a journal for computing internationality. Internationality, y is defined as a multivariate function of $x_i$, i=1, 2...n. Internationality score varies over time and depends on scholastic

The scibase and scientometric modeling effort is endorsed and supported by IEEE Computer Society Bangalore Chapter.

parameters, subject to evaluations, constant scrutiny and ever changing patterns.

## 2. Cobb Douglas Model

In economics, Cobb-Douglas production function [2,5,6] is widely used to represent relationship of outputs to inputs. This is a technical relation which describes the Laws of Proportion, i.e., the transformation of factor inputs into outputs at any particular time period. This production function is used for the first time, to compute the internationality [3] of a journal where the predictor/independent variables, $x_i$, i=1, 2…n are algorithmically extracted from curated data.

Cobb-Douglas function is given by:

$$y = A\prod_{i=1}^{n} x_i^{\alpha_i}$$

Where, $y$ is the internationality score, $x_i$ are the predictor variables/input parameters and $\alpha_i$ are the elasticity coefficients. The function has extremely useful properties such as convexity/concavity depending upon the elasticities. The properties yield global extrema which are intended to be exploited in the computation of internationality or influence. For $n = 4$, $x1$ to $x4$ are the input parameters as described below.

  $x1$:other-citations quotient)=[1-(self-citations /Total citations)]
  $x2$ : International Collaboration
  $x3$ : Source Normalized Impact per paper (SNIP)
$x4$ : Non-Local Influence Quotient = [1 -(Journal's self-citations /Total citations)]

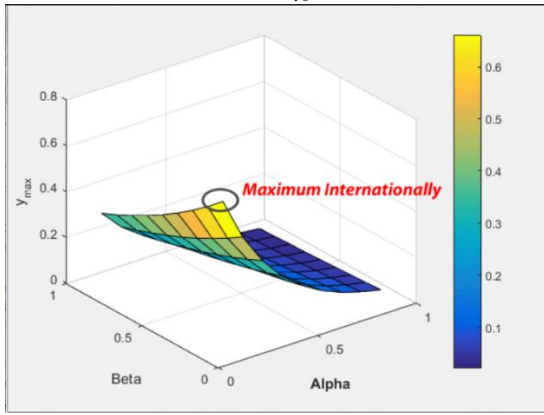

Figure2. For optimum values of elasticity, the Cobb Douglas function attains a maximum value.

## 3. Graph data modeling

Data accumulated is rich, very well connected and has a lot of hidden information within it. Hence, we chose to visualize this data using graph database. A graph database is a graph-oriented database, which is type of NoSQL database that uses graph theory to store, map and query relationships. It is basically a collection of nodes and edges [1, 7]. Graph data modeling [8] is the procedure in which a Neo4j user depicts a subjective space as a connected graph of nodes and relationships. From this description, a graph data model is designed to answer questions in the form of Cypher queries.

Scholarly articles and scientific journals make the most of our research area hence we identified the elements from these which can then be transformed into nodes and relationships. Few of the elements constitute properties as shown in the square boxes beside the oval shaped nodes. Fig 3 is the data model that has been designed using the Neo4j graph data modelling.

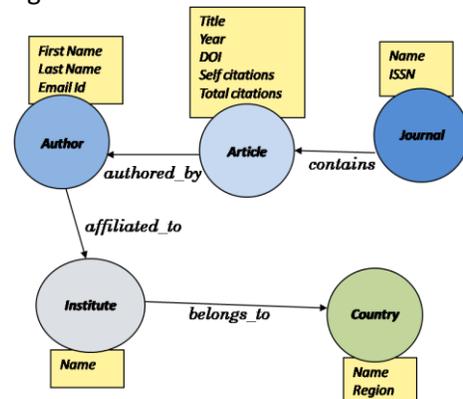

Fig 3 Data model for Graph database, the model depicts the overall structure of accumulated data. It shows relationship between journal, articles published in that journal, contributing authors, their affiliation and the country.

## 4. Data visualization

Scraped data is imported into graph database of Neo4j. Few of the visualizations that are possible with our data model are:
- Author network
- Institute network
- Country network
- Spread of a domain in a country
- Collaboration network of an institute
- Extract year-wise publication trend of an author

This data when queried appropriately can help to visualize the shape of 'Internationality' of a Journal at various levels. Few of the visualization are as following.

The scibase and scientometric modeling effort is endorsed and supported by IEEE Computer Society Bangalore Chapter.

### a. Journal to Author to Country mapping

Figure 4, shows mapping of a journal to Author and to Country, to which Institution belongs. The blue circles represent journal nodes, purple circles are author nodes, yellow represent article nodes and red ones are country nodes. These links will help in identifying degree of contribution of countries and regions to a domain. Since we can identify spurious journals using journal's internationality modeling index (JIMI), we can now identify which regions are essentially contributing more to such nexus of fake and dubious journals.

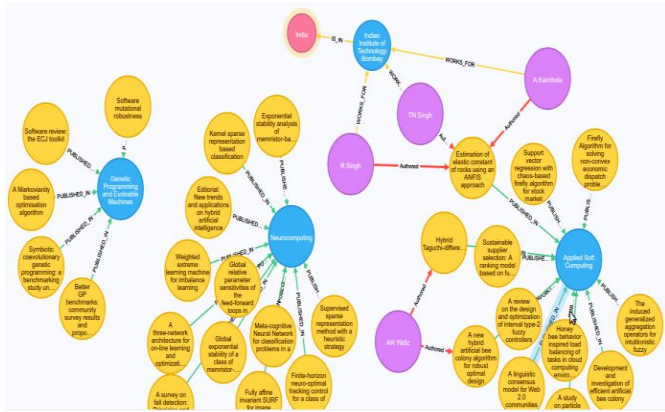

Figure 4. Journal to Author to Country mapping: red colored node represents country (India), purple node are authors, yellow nodes are title of the articles they authored and blue nodes are affiliating institutes.

### b. Author to Institute to Country mapping

Figure 5 shows mapping of the Author to Institute to Country. Blue nodes represent Institution, red represent country and purple represent author nodes. This mapping can help in identifying the contribution trends pertaining to a particular Institute and Country in particular.

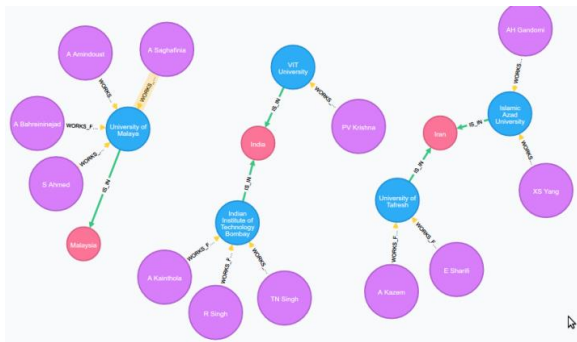

Figure 5. Author to Institute to Country mapping: red nodes are the countries, blue nodes are institutes and purple nodes shows authors who are serving these institutes

### c. Article to Author Mapping

Figure 6 is visualization of a particular author's contribution in totality. We can query this data based on year to visualize year by year contributions made. In the figure purple nodes are authors and yellow nodes are articles. For a rich and dense data we can provide visually appealing information about an author's reach and contributions using this visualization.

### d. Institute to Country to Region mapping

Figure 7 is the data for all of the institutes belonging to a country, and countries belonging to a Region. These links can help in identifying contribution of various institutes and respective countries to a domain

All the articles with a degree of Article impact, when measured using Singular value Decomposition for that domain and the journals with a measure of 'internationality' (computed using Cobb Douglas model) will define the degree of contribution made by any Institute and in turn Country to a particular scientific domain (to which that journal belongs to).

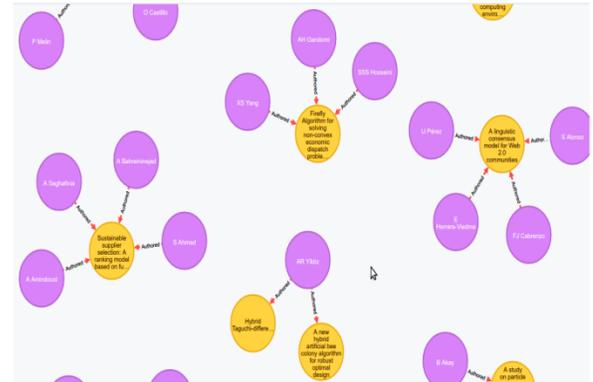

Figure 6 Author to Article mapping: purple nodes depicts authors and yellow nodes are published articles

In other words, these two measures i.e. Article impact and the journal's 'internationality' index can be used to define contribution of any Author, Institute, Country and Region to a particular domain.

The scibase and scientometric modeling effort is endorsed and supported by IEEE Computer Society Bangalore Chapter.

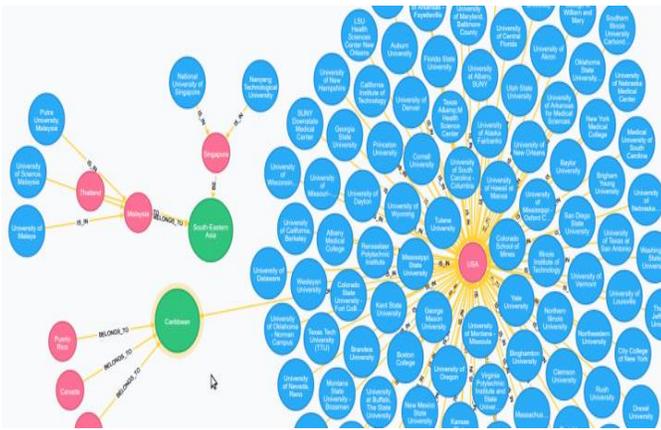

Figure 7. Institute to Country to Region mapping: red nodes are the countries, blue nodes are institutes and green nodes are regions. The figure shows that US has large number of institutes.

Conversely these two parameters can be used as a scale to evaluate Authors, journals, Institutes, Countries and Regions. This scale can explain the growth and contribution made by all of these. When the data is large enough we can predict evolving field, most dominating country in a particular field/domain and increase or decrease of impact for any given journal.

**RESULTS**

Following are the various cypher queries and resulting visualization from the graph database. Fig 8 depicts journal to Author and in turn to the country of affiliation mapping. Following is the query on the graph database to extract the needed information for graph plotting.

**Cypher Query:**
MATCH (Journal)<-[:PUBLISHED_IN]-(Article) WHERE Journal.name IN ['Applied Soft Computing', 'Neurocomputing', 'Genetic Programming and Evolvable Machines'] RETURN Article.year, Journal.name

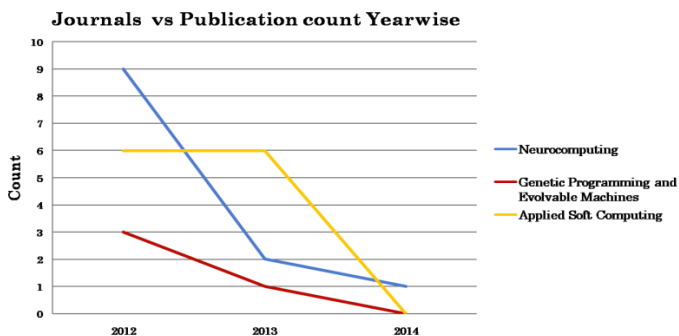

Figure 8. Line graph; Journal Vs Publication

Fig 9 shows the area graph of total citations vs. self-citations for all the articles and journals in the graph database. Following is the query which extracts this data.

**Cypher query:**
MATCH (n:Article) RETURN n.totalcites, n.selfcites

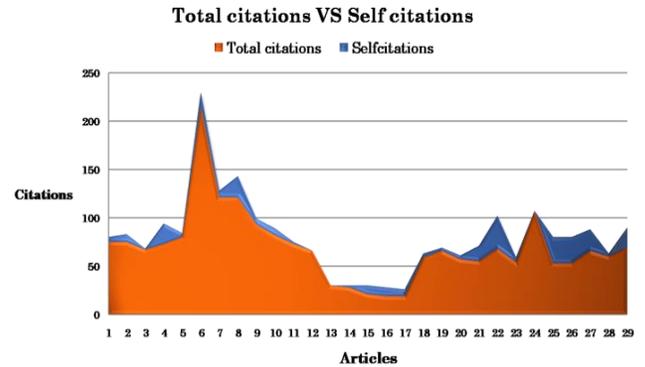

Figure 9. Area graph; Total citation Vs. Self-citations

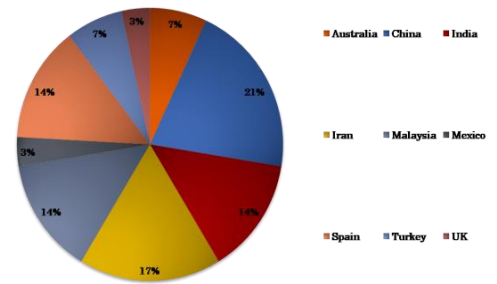

Figure 10. Pie graph; Article Publications per Country

Fig 10 shows the Pi graph of Article Publications per Country. Following is the query which is executed on the database to extract this data.

**Cypher Query:**
MATCH (Author)-[r:WORKS_FOR]->(Institute)-[s:IS_IN]->(Country) RETURN Author.name, Country.name

**SCIBASE**

SciBase[7] is a project started in 2015 with an aim to collect and store information on journals, authors and articles to facilitate the data on scholastic indicators to the emerging and established researchers across the globe who work on scientometrics and related domains. The database is a web dictionary, which provides information on journals, articles published by journals, contributing

The scibase and scientometric modeling effort is endorsed and supported by IEEE Computer Society Bangalore Chapter.

authors and many related information like author's affiliation, journal's country information etc. The data is mainly for ACM and IEEE journals. The repository is built by running python scripts to scrape web pages and later, storing information in JSON format. The scholastic information is provided in CSV and JSON format both in downloadable form and graphical representation of the data is also shown for quick understanding and interpretation. SciBase assimilate 3 prominent features - VizKit, RREF and OPRS. VizKit (Visualization Kit) enables its user to explore hidden patterns in journal/author/article data in visual manner. The visual effect speeds up the process of understanding and deriving newer metrics from scholastic data. RREF, also known as recursive referencing of articles, depicts recursive representation of an article's reference list. The repository shows the RREF graph for articles of Dr. Terrence Tao. OPRS, Open peer-review system, is a platform that allows researchers to submit articles for open peer-review.

The visualization package shown above is integral to SciBase project which helps in deriving information on how widespread is the diffusion of influence with respect to authors, institutes, and countries and also with respect to journals. One would like to know the number of articles published in a domain by a set of authors belonging to certain institutes and correspondingly, it may be interesting to know the institutes or countries that are leading in terms of research progress. Figures 4, 5, and 6 depict visualizations that highlight association of authors (majorly within the institute they are affiliated to) who have contributed remarkably in a specific domain and also bring those journals into focus, that are preferred by these authors for publications. Figure 4, journal to author mapping shows author preferences for particular journals. Alongside, the figures bring out the degree of influence-diffusion of various journals by comparing the country information of contributing authors with that of the journals. Larger the number of published articles that came through authors of varied countries, greater is the influence diffused by the publishing journal in that domain. Institute-wise, Journal to author to institute mapping shows profile of institute publishing in high impact journals and draw particular attention to institutes that are not publishing in high impact journals. With all this indicators, SciBase derives crucial parameters, one of which is NLIQ (Non-Local Influence Quotient). NLIQ emphasize the spread of influence of authors, institutes and journals thereby help identifying leaders in respective domains.

## CONCLUSION

Scholarly articles and scientific journal datasets need special type of database as the data is massive in scale and ever evolving. Various web-scraping and parsing techniques were used to create and develop a platform for ScientoBASE [7], a repository, which will consist of international journals by subject category with ranks and scores of internationality and necessary metric information. Graph database such as Neo4j, which is a NoSQL type of database is an emerging technology in the field of effective visualization and data storage. It not only provides a more meaningful method of data storage but also facilitates intelligent query formulation for the meaningful data extraction and analysis. In this research Neo4j has played a crucial role in finding the hidden patterns which can further enhance the usability of the information concealed within the huge databases maintained worldwide. This work will lead to software which will be an end-to-end product comparable with Scopus and ISI's Web of Science but positioned in a distinct space and cater to the needs of the underprivileged researchers in developing countries.

The extensive point of our exploration is to characterize a yardstick of scientific contribution and international diffusion; especially in niche areas such as Astroinformatics, Computational Neuroscience, Industrial Mathematics and Data Science from India, as well as other countries across the globe. The result of our examination will clear path for information and model approval and development of an information perception and web interface apparatus that will compute the scores and provide visualizations of every vital parameter of internationality. This tool can be used as a web toolkit to quantify the growth of Indian as well as worldwide Scientometry in cutting edge and rising territories in Science and Technology. A crucial aim of the SciBase project is to be able to exploit hidden patterns and relationships in the SciBase Graph to derive metrics for deeper understanding and analysis of author/article/journal data. The VizKit platform allows users to do just that albeit in an elegantly visual manner. Beta version of the desktop application is available for download at http://sahascibase.org/vizkit

The scibase and scientometric modeling effort is endorsed and supported by IEEE Computer Society Bangalore Chapter.

The scibase and scientometric modeling effort is endorsed and supported by IEEE Computer Society Bangalore Chapter.

The scibase and scientometric modeling effort is endorsed and supported by IEEE Computer Society Bangalore Chapter.